\documentclass[12pt,superscriptaddress,reprint]{revtex4-1}

\usepackage[usenames]{color}
\usepackage{hyperref}

\usepackage[dvipsnames]{xcolor}

\setcitestyle{super}

\hypersetup{ 
colorlinks=true,
linkcolor=blue,
citecolor=blue
}

\newcommand{\linecite}[1]{\hspace{-1 ex} \nocite{#1}\citenum{#1}}

\usepackage{lineno}
\usepackage{graphicx}

\begin{document}

\author{Brett Leedahl}
\affiliation{Max Planck Institute for Chemical Physics of Solids, N{\"o}thnitzer Stra{\ss}e 40, 01187 Dresden, Germany}

\author{Martin Sundermann}
\affiliation{Max Planck Institute for Chemical Physics of Solids, N{\"o}thnitzer Stra{\ss}e 40, 01187 Dresden, Germany}
\affiliation{Institute of Physics II, University of Cologne, Z\"{u}lpicher Stra{\ss}e 77, D-50937 Cologne, Germany}

\author{Andrea Amorese}
\affiliation{Max Planck Institute for Chemical Physics of Solids, N{\"o}thnitzer Stra{\ss}e 40, 01187 Dresden, Germany}
\affiliation{Institute of Physics II, University of Cologne, Z\"{u}lpicher Stra{\ss}e 77, D-50937 Cologne, Germany}

\author{Andrea Severing}
\affiliation{Max Planck Institute for Chemical Physics of Solids, N{\"o}thnitzer Stra{\ss}e 40, 01187 Dresden, Germany}
\affiliation{Institute of Physics II, University of Cologne, Z\"{u}lpicher Stra{\ss}e 77, D-50937 Cologne, Germany}

\author{Hlynur Gretarsson}
\affiliation{Max Planck Institute for Chemical Physics of Solids, N{\"o}thnitzer Stra{\ss}e 40, 01187 Dresden, Germany}
\affiliation{PETRA III, Deutsches Elektronen-Synchrotron (DESY), Notkestra{\ss}e 85, 22607 Hamburg, Germany}

\author{Lunyong Zhang}
\affiliation{Max Planck Institute for Chemical Physics of Solids, N{\"o}thnitzer Stra{\ss}e 40, 01187 Dresden, Germany}

\author{Alexander C. Komarek}
\affiliation{Max Planck Institute for Chemical Physics of Solids, N{\"o}thnitzer Stra{\ss}e 40, 01187 Dresden, Germany}

\author{Maurits W. Haverkort}
\affiliation{Institute for Theoretical Physics, Heidelberg University, Philosophenweg 19, 69120 Heidelberg, Germany}

\author{Antoine Maignan}
\affiliation{Laboratoire CRISMAT, UMR 6508 CNRS-ENSICAEN, 6 bd Mar\'echal Juin, 14050 Caen Cedex, France}

\author{Liu Hao Tjeng}
\affiliation{Max Planck Institute for Chemical Physics of Solids, N{\"o}thnitzer Stra{\ss}e 40, 01187 Dresden, Germany}
\email{hao.tjeng@cpfs.mpg.de}

\title{Origin of Ising magnetism in Ca$_3$Co$_2$O$_6$ unveiled by orbital imaging}

\newpage

\begin{abstract}
The one-dimensional cobaltate Ca$_3$Co$_2$O$_6$ is an intriguing material having an unconventional magnetic structure, displaying quantum tunneling phenomena in its magnetization. Using a newly developed experimental method, $s$-core-level non-resonant inelastic x-ray scattering ($s$-NIXS), we were able to image the atomic Co $3d$ orbital that is responsible for the Ising magnetism in this system. We show that we can directly observe that it is the complex $d_2$ orbital occupied by the sixth electron at the high-spin Co$_{\text{trig}}^{3+}$ ($d^6$) sites that generates this behavior. This is extremely rare in the research field of transition metal compounds, and is only made possible by the delicately balanced prismatic trigonal coordination. The ability to directly relate the orbital occupation with the local crystal structure is essential to model the magnetic properties of this system.
\end{abstract}

\maketitle

\newpage

{\raggedleft\textbf{Introduction}}\\
Since its crystal structure was fully determined in 1996\cite{Fjellvag1996}, Ca$_3$Co$_2$O$_6$ has garnered a large degree of attention due to its special atomic arrangements and peculiar magnetic properties. The discovery of stair-step jumps in the magnetization at regular intervals with an increasing applied field strength \cite{Aasland1997,Kageyama1997,Maignan2000,Hardy2004} is indicative of the presence of quantum tunneling phenomena \cite{Maignan2004}. This has triggered a flurry of theoretical and experimental research activities on Ca$_3$Co$_2$O$_6$ \cite{Vidya2003,Wu2005,Kudasov2006,Agrestini2008,Agrestini2011,Kamiya2012,Kozlenko2018,Maignan2004} and its close derivatives \cite{Niitaka2001,Sugiyama2006,Choi2008,Wu2009}.

\begin{figure}[!h]
\begin{center}
\includegraphics[width=3.375in]{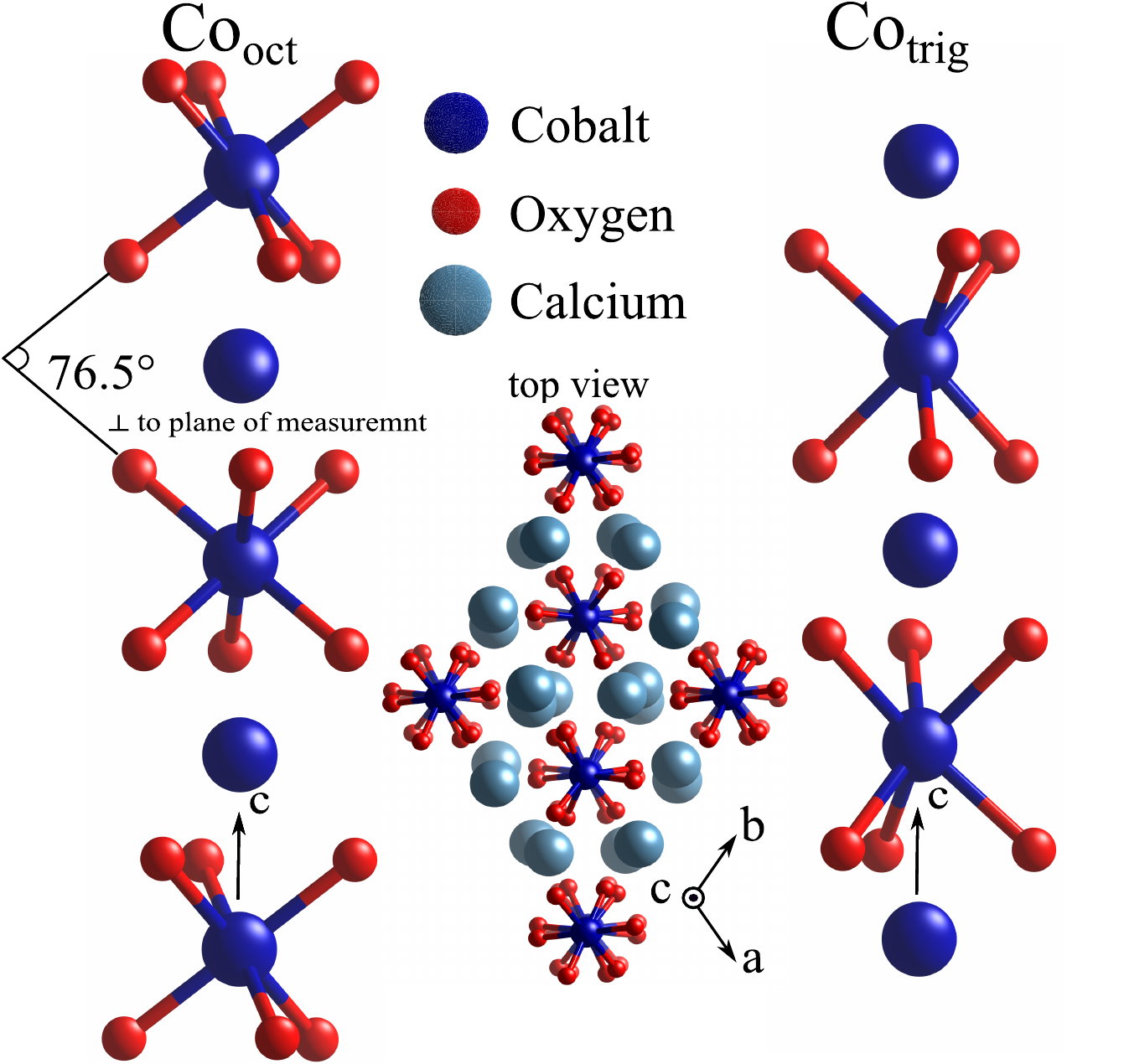}
\caption{\textbf{Ca$_3$Co$_2$O$_6$ crystal structure} The crystal 
structure of Ca$_3$Co$_2$O$_6$ consists of chains of Co atoms along the $c$-axis, where each 
cobalt site alternates between trigonal and octahedral coordination with the surrounding oxygen 
ligands. Additionally, both trigonal and octahedral sites alternate, with each having a 45 degree rotation about the $c$-axis relative to their nearest neighbor (along the $c$-axis) of the same coordination. In the $ab$-plane (top view), the chains form a triangular type lattice}
\label{fig:struct}
\end{center}
\end{figure}

One-dimensional chains are formed in Ca$_3$Co$_2$O$_6$  along the $c$-axis with alternating CoO$_6$ octahedra and trigonal prisms. In the $ab$-plane, the chains form a triangular type lattice\cite{Fjellvag1996} (see Fig. \ref{fig:struct}). The intra-chain  Co--Co coupling is ferromagnetic, while the inter-chain coupling is antiferromagnetic \cite{Aasland1997}. The unusual magnetic properties, along with its Ising-like character\cite{Maignan2000,Hardy2003,Maignan2004}, are linked to the geometrically frustrated crystal lattice.

However, the origin or character of the presumed Ising magnetism is the subject of much discussion. To resolve this, one actually has to also simultaneously address the issue of the charge states of each of the two Co sites, as well as their spin-states \cite{Vidya2003,Whangbo2003,Eyert2004,Sampathkumaran2004,Takubo2005,Wu2005,Burnus2006}. All of the previous studies relied on calculations of one type or another, with varying and conflicting outcomes, depending on what theoretical approach was used.

Herein, we make use of $3s$ core level non-resonant inelastic scattering (NIXS) to determine the orbital occupation of the cobalt $3d$ ions. As has recently been shown, the angular dependence of the NIXS integrated intensity of the $3s$\,$\rightarrow$\,$3d$ transition can directly map out the shape of the local orbital hole density of an ion in its ground state, without the need to do any calculational modelling of the spectral line shapes\cite{Yavas2019}. Applying this novel technique to our system, we are able to identify the Co $3d$ orbital that is at the core of the Ising behavior. Moreover, from the experimentally determined orbital occupations we are able to unequivocally establish both the charge and spin state of each of the two types of Co ions in the crystal. \\

\begin{figure}
\begin{center}
\includegraphics[width=3.375in]{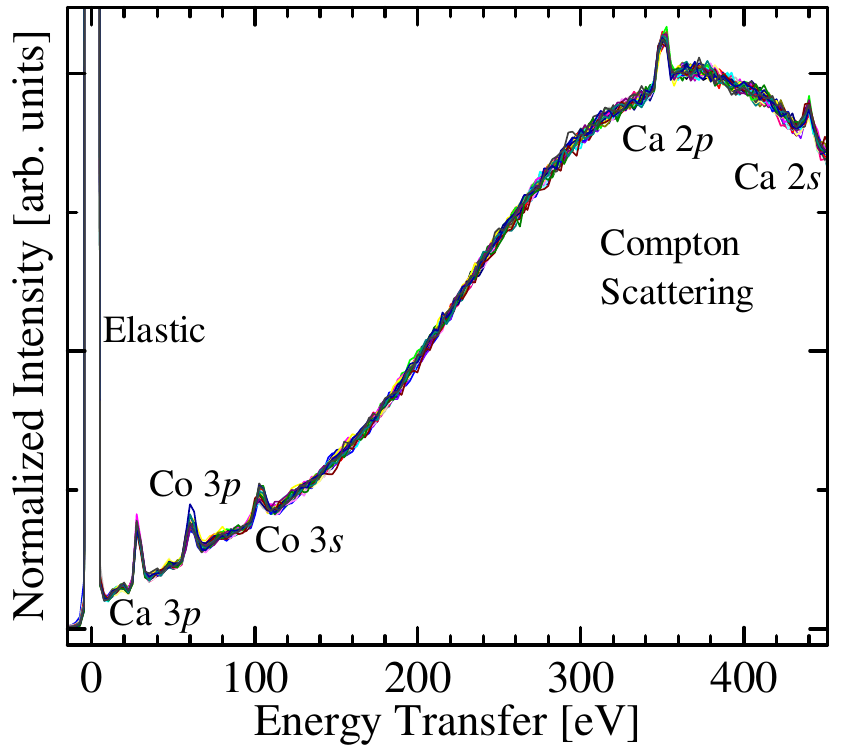}
\caption{\textbf{Survey spectra of scans from all sample rotation angles.} The energy was scanned from the elastic peak (9.69\,keV) to 450\,eV above it, in order to observe the Compton scattering profile used for normalization. All calcium and cobalt absorption edges with core-level electron binding energies in the range can readily be seen. Of note is that dipole forbidden Co $3s$\,$\rightarrow$\,$3d$ transitions are visible.}
\label{fig:compton}
\end{center}
\end{figure}

\begin{figure}
\begin{center}
\includegraphics[width=3.375in]{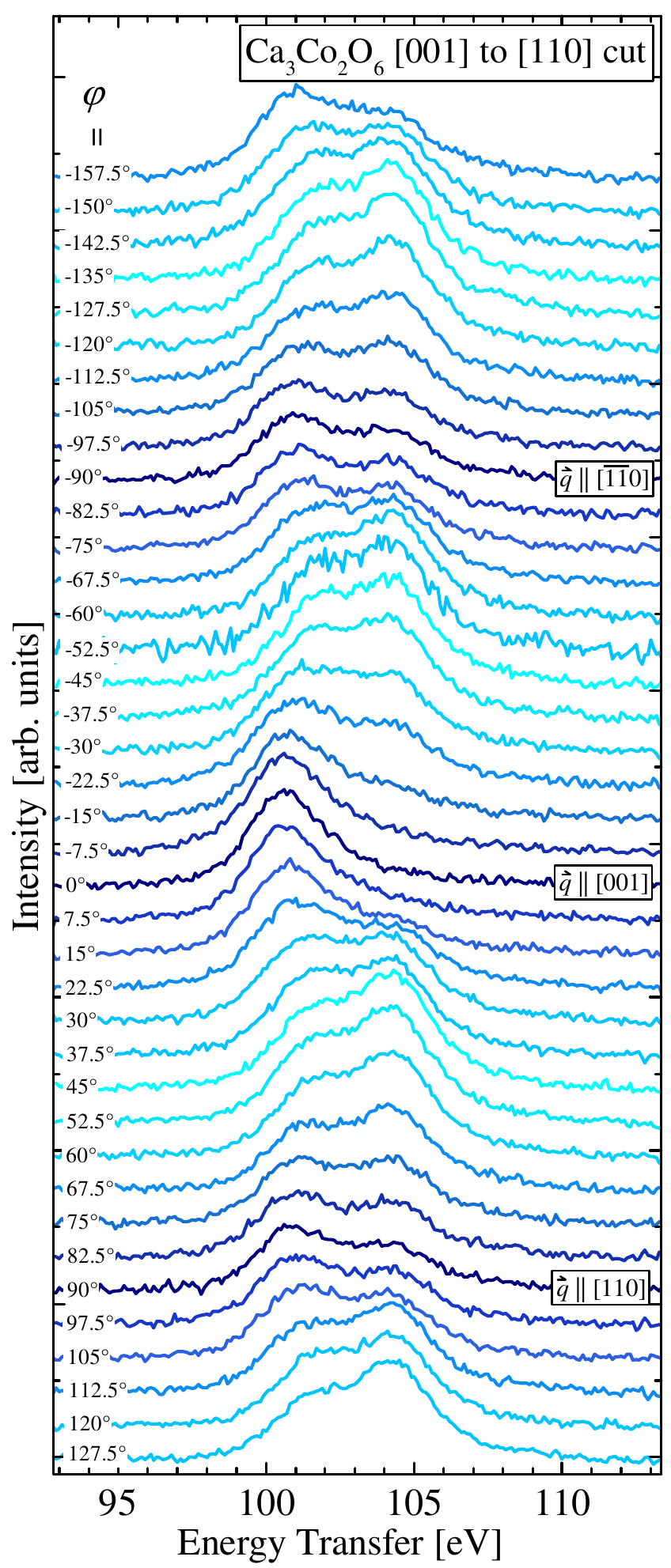}
\caption{\textbf{NIXS spectra at the Co $M_1$ edge.} An array of spectra were recorded by rotating the sample in 7.5$^\circ$ increments such that $\vec{q}$ remained in the [110] to [001] crystallographic plane. The integrated intensity of the peaks is directly proportional to the hole density of the Co $d$-electrons in the direction of the momentum transfer $\vec{q}$. The color coding is such that the darkest blue curves are spectra taken with $\vec{q}$ parallel to high symmetry directions of the single crystal Ca$_3$Co$_2$O$_6$.}
\label{fig:stackedspectra}
\end{center}
\end{figure}

{\raggedleft\textbf{Results}}

Our experimental setup is illustrated in Ref. \linecite{Yavas2019} and a general description of the inelastic x-ray scattering method is given in Ref. \linecite{Sahle2015}. Incoming x-rays ($\sim$\,10keV) were scattered by the Ca$_3$Co$_2$O$_6$ single crystal with momentum transfer $\vec{q}=\vec{k_i}-\vec{k_f}$ and energy transfer $\hbar\omega=\hbar\omega_i-\hbar\omega_f$, where $\vec{k}_{i,f}$ and $\hbar\omega_{i,f}$ denote the momentum and energy of the incoming and scattered photons, respectively (see Methods). We recorded the scattered beam as a function of the sample angle $\varphi$, here defined as the angle between the fixed momentum transfer vector $\vec{q}$ and the $c$-axis of the crystal. 

A collection of NIXS spectra measured at various sample angles is displayed in Fig. \ref{fig:compton}, all of which have been normalized to have the same maximum intensity at $\approx$\,370\,eV. The spectra show the $M_{2,3}$ edges ($3p$\,$\rightarrow$\,$3d$) of calcium at $\approx$\,30\,eV and cobalt at $\approx$\,60\,eV and, most importantly, the dipole-forbidden $M_1$ ($3s$\,$\rightarrow$\,$3d$) excitations of cobalt at around $\approx$\,100\,eV, all overlaid on a broad Compton scattering profile---caused by the inelastic scattering of photons off the electron density of the material. A close-up of the cobalt $M_1$ edge and its angular dependence on $\varphi$ is displayed in Fig. \ref{fig:stackedspectra}. Each spectrum was measured at a different sample angle $\varphi$ in the crystallographic [001] to [110] plane, with a 7.5$^\circ$ increment between successive spectra. We have labeled  $\vec{q}$\,$\parallel$\,$[110]$, $\vec{q}$\,$\parallel$\,$[001]$, and $\vec{q}$\,$\parallel$\,$[\overline{11}0]$ for measurements in which the momentum transfer vector was parallel to a high symmetry direction of the crystal. Note that the Compton profile in the narrow $M_1$ region has been subtracted using a  linear background (see Methods).

\begin{figure}
\begin{center}
\includegraphics[width=3.375in]{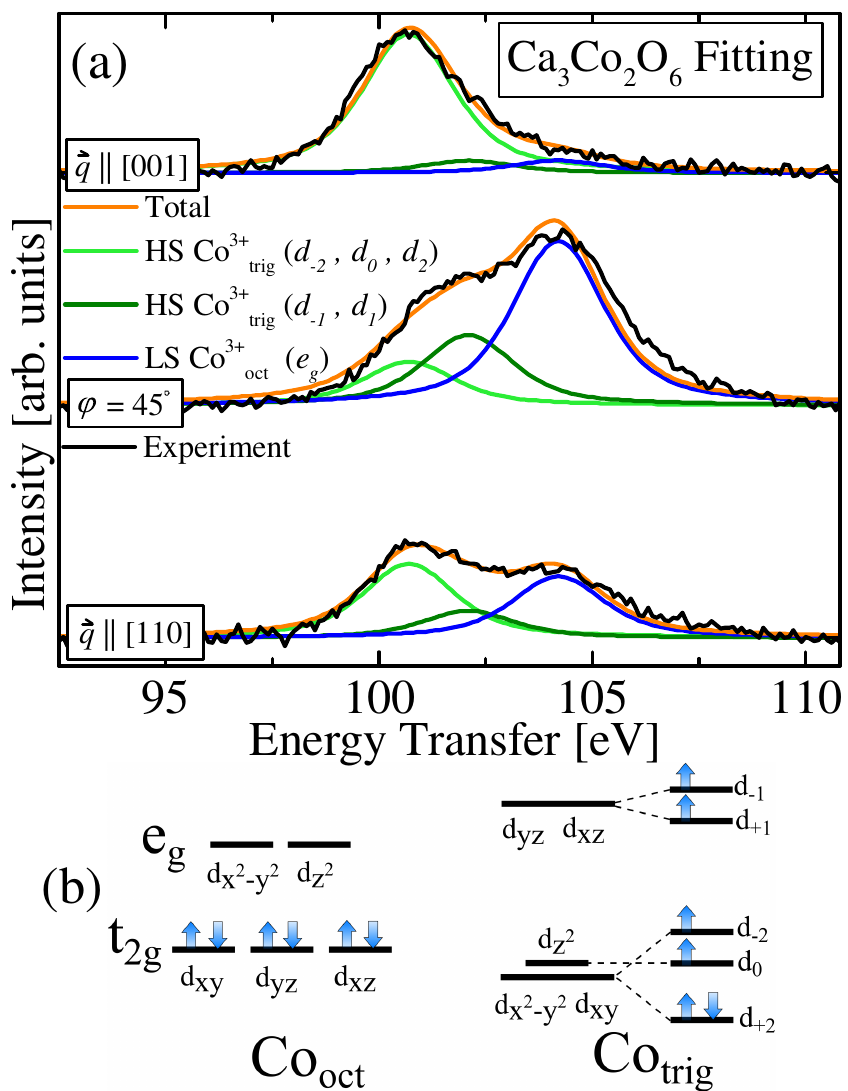}
\caption{\textbf{Voigt function fitting of experimental NIXS spectra.} (a) The experimental spectra (black) 
were fit with Voigt functions that are a convolution between a 1.4\,eV FWHM Gaussian function (to model 
the experimental resolution), and a 1.8\,eV FWHM Lorentzian function (to model lifetime broadening). 
The component curves were integrated for each spectrum of Fig. \ref{fig:stackedspectra}, and this value is plotted on the polar plot of Fig. \ref{fig:110to001}, as a function of sample angle. (b) A schematic energy level diagram for the Co$_{\text{oct}}$ site in a 3+ low spin state shows that all the holes are in the $e_g$ orbitals; while for the Co$_{\text{trig}}$ in a 3+ high spin configuratioin, the sixth and minority-spin electron occupies the $d_2$ orbital.}
\label{fig:fitting}
\end{center}
\end{figure}

The spectra in Fig. \ref{fig:stackedspectra} taken at different sample angles in the [001] to [110] plane are composed of several peaks. We show a selection of these spectra in Fig. \ref{fig:fitting}(a) to demonstrate that we can discern three peaks, one at 100.7\,eV, one at 104.2\,eV, and one in-between at 102.1\,eV. Once given our final results, one will see that there is good \emph{a posteriori} justification for the presence of these three peaks, and the electron configurations in Fig. \ref{fig:fitting}(b). 

To quantitatively analyze the spectra of Fig. \ref{fig:stackedspectra}, we decomposed each spectrum into a linear combination of these three peaks, for which we are interested in their individual intensities. As illustrated in Fig. \ref{fig:fitting}(a), we carried out a fitting procedure in which the peaks were fit with Voigt functions to model the 1.4\,eV FWHM Gaussian experimental energy resolution, and the lifetime broadening, for which a 1.8\,eV FWHM Lorentzian provides the optimal fit. When implementing the fitting procedure across the array of spectra of Fig. \ref{fig:stackedspectra}, the Gaussian and Lorentzian widths of the Voigt functions were fixed, as well as the peaks' central energy positions throughout; only the intensities were free to vary to obtain the best fit to the experimental spectra. 

\begin{figure*}
\begin{center}
\includegraphics[width=6.5in]{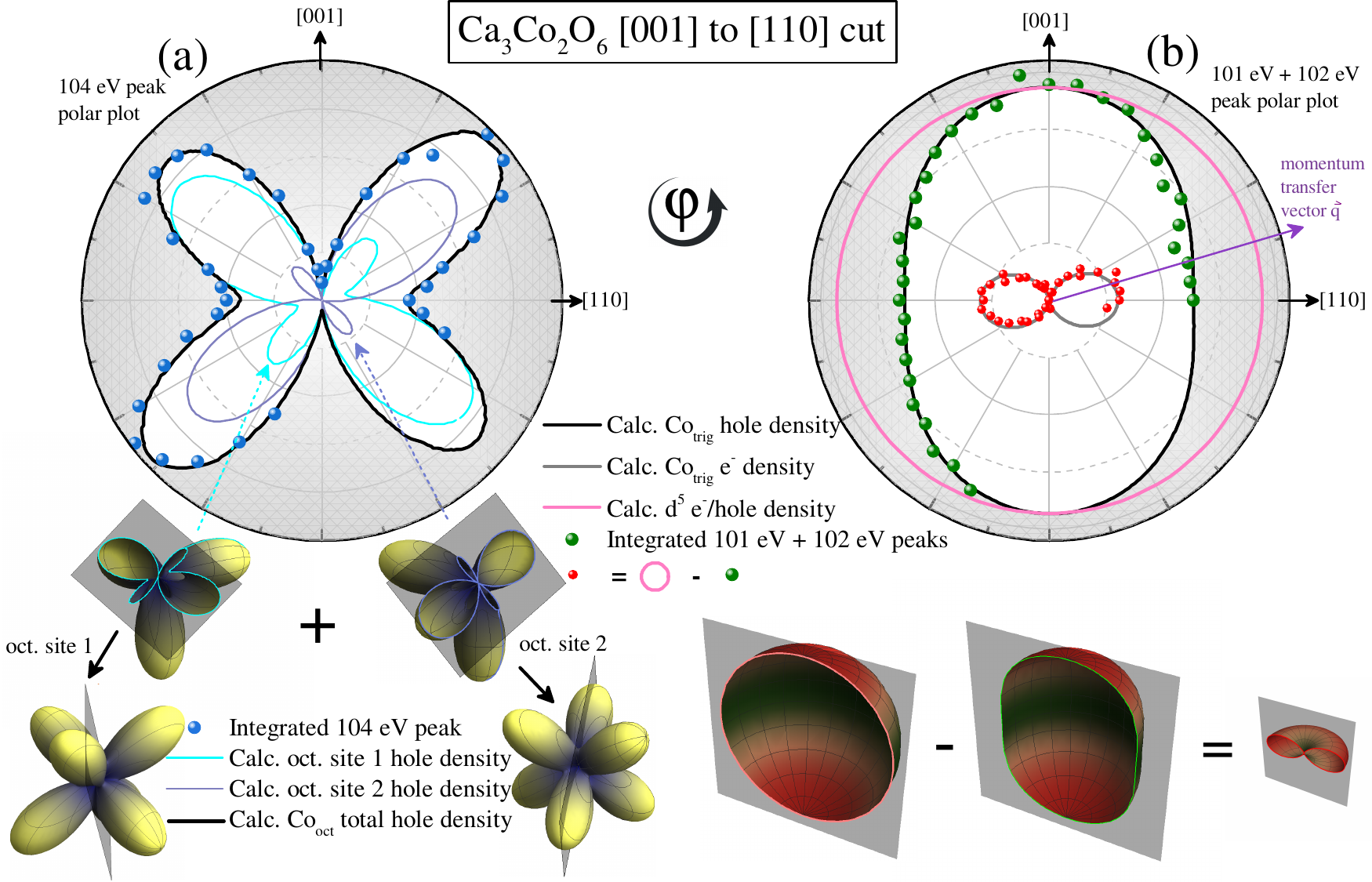}
\caption{\textbf{Integrated intensities of the Co $M_1$ ($3s$\,$\rightarrow$\,$3d$) spectra of Fig. \ref{fig:stackedspectra}.} The data points (derived from integrated intensities) are plotted as a function of the angle $\varphi$ between the crystallographic [001] axis and the momentum transfer vector $\vec{q}$. The 3D images highlight the relevant cuts taken in the experiment in the context of the total $e^-$/hole densities. (a) The blue dots are the result of integrating the 104.2\,eV peak of Figs. \ref{fig:stackedspectra} and \ref{fig:fitting}. They agree very well with the calculated total Co$_{\text{oct}}$ hole density (black) for the candidate $d^6$ LS configuration, which is a sum of the two Co$_{\text{oct}}$ sites' hole densities (blue solid lines). The blue solid lines in the polar plot correspond to the cross-sectional cuts shown in the 3D figures beneath. (b) The calculated hole density for a $d^6$ HS configuration wherein the sixth electron occupies the complex $d_{2}$ orbital is shown in black, which agrees very well with our data points derived from the integrated 100.7\,eV + 102.1\,eV peaks. We can obtain the orbital wavefunction of the sixth electron (both experimentally and theoretically) by subtracting the hole density (green and black) from a reference $d^5$ circular hole density (pink). This result is shown experimentally by the red dots and theoretically with the grey solid line. The agreement reveals that the Ising-like magnetism originates from the sixth electron, which occupies the $d_2$ orbital.}
\label{fig:110to001}
\end{center}
\end{figure*}

To properly appreciate the intensity variations in the spectra one must map them to a polar plot. This is shown in Fig. \ref{fig:110to001}, which displays the results from the angular scans in the [001] to [110] plane. The integrated intensity of the fitted peaks are plotted as a function of the measurement angle $\varphi$. The blue dots in Fig. \ref{fig:110to001}(a) show the angular dependence of the intensity of the 104.2\,eV peak, while the green dots in Fig. \ref{fig:110to001}(b) show that of the summed intensities of the 100.7\,eV and 102.1\,eV peaks. 

To interpret the four-lobe shape of the 104.2\,eV polar plot (Fig. \ref{fig:110to001}(a)), we start with the \emph{ansatz} that the Co$_{\text{oct}}$ ion is 3+ and low-spin (LS), i.e. $3d^6$ with a $t_{2g}^6$ configuration. This implies that all four holes of this ion will be in the $e_g$ shell, as depicted in Fig. \ref{fig:fitting}(b).  To explain the asymmetries in the four-lobe shape, it is necessary to consider some of the more subtle details of the crystal structure. There are three details regarding the local coordination of Co$_{\text{oct}}$ sites that concern us: (1) along the Co--Co chains there are alternating Co$_{\text{oct}}$ sites that are rotated 45 degrees about the $c$-axis from one another; (2) the axes along which the $d_{z^2}$ lobes lay for the two sites are oriented 13.5 degrees off right angles from one another. Both of these are apparent in Fig. \ref{fig:struct}. Lastly, (3) it is also necessary to consider the effect of the crystal field, which deviates slightly from that of nominal $O_h$ symmetry because the bond angles of the octahedral sites are 3 degrees off from the nominal 90 degrees.

What follows, is that due to (1), the experimental cut is a sum of a slice through the large $e_g$ lobes of one site, plus the large $e_g$ lobes of the other site. These two octahedral sites, along with the plane of the cut taken in the experiment are illustrated by the two lower 3D blue-yellow shapes in Fig. \ref{fig:110to001}, which show identical planes, with only the sites rotated 45 degrees from one another as in the real crystal. In regards to (2), this means that the sum of the two sites is not done at right angles to one another, but at 76.5 degrees. This is evident in the polar plot of Fig. \ref{fig:110to001}(a), which shows the contributions from both sites as solid blue lines of different shades. Finally, (3) has the more subtle effect of slightly altering the relative lobe intensities. Taking into account all these details of the crystal structure, the theoretically expected curve was calculated, and is shown as a black solid line in Fig. \ref{fig:110to001}(a)---which is a sum of the two blue curves. The near perfect agreement between the experimental data points and the calculated curve establishes that the \emph{ansatz} is correct, i.e. that the  Co$_{\text{oct}}$ ions are 3+ and LS, which suggests that the magnetism must originate from the trigonal site. 

\begin{figure}
\begin{center}
\includegraphics[width=3.375in]{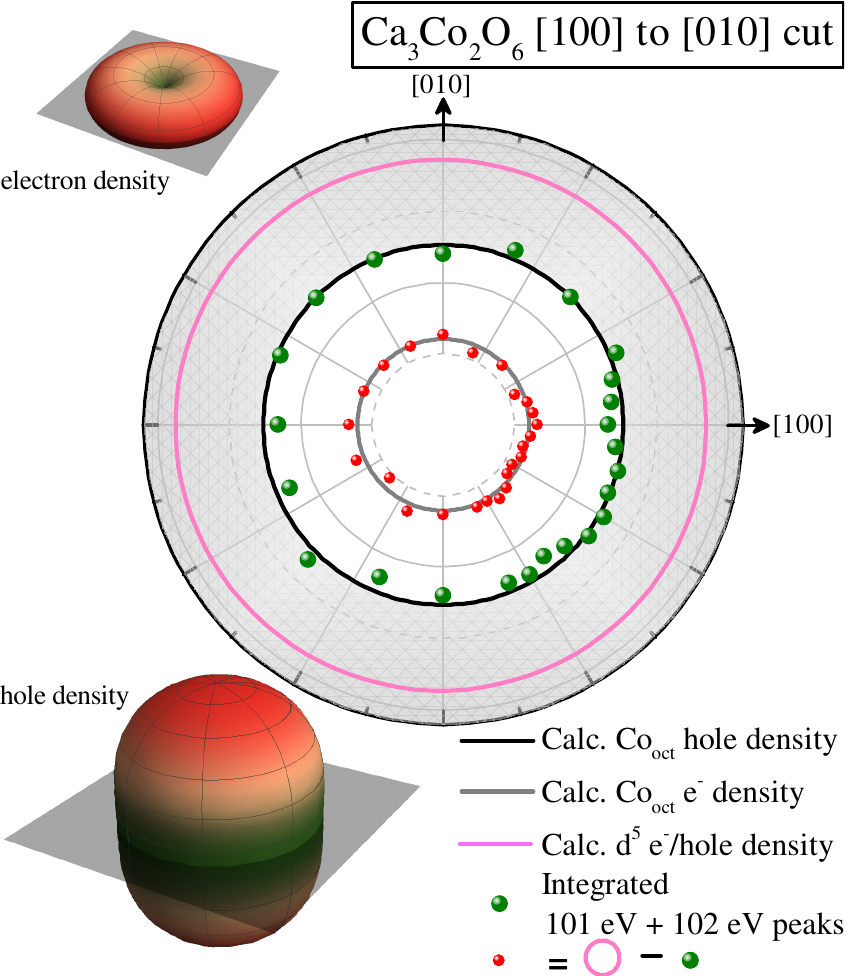}
\caption{\textbf{The Co$_{\text{trig}}$ site electron density cut in the XY-plane.} The experimental/theoretical electron density (red/grey) was obtained by subtracting the experimentally/theoretically determined hole density (green/black) from a reference $d^5$ circular (spherical in 3D) hole density (pink), which is of the same radius as in Fig. \ref{fig:110to001}. The insets show the calculated 3D shapes of the electorn and hole densities, along with the transparent grey planes that correspond to the experimental [100] to [011] cut. Again, we see that the circular shape of the $d_2$ orbital in this plane agrees well with our experimental data points.}
\label{fig:100to010}
\end{center}
\end{figure}

We now focus our attention on the remaining intensity not yet accounted for, that is, the 100.7\,eV and 
102.1\,eV peaks of Fig. \ref{fig:fitting}, which must result from excitations at the Co$_{\text{trig}}$ site. Given that the Co$_{\text{oct}}$ is 3+, charge balance requires that the Co$_{\text{trig}}$ ions must also be 3+. The total integral of the 100.7\,eV and 102.1\,eV peaks leads to the conclusion that the hole density of the Co$_{\text{trig}}$ $3d^6$ ion has an oval shape, as shown by the green dots in Figure \ref{fig:110to001}(b). To understand the meaning of this shape, consider first a high-spin (HS) $3d^5$ ion. Its hole density will be spherical since it is a half-filled shell system, where each of the five $d$-orbitals contains one electron. To represent this situation in the [001] to [110] plane of our experiment, we have added a pink circle to Figure \ref{fig:110to001}(b). The difference between the circle (reference $d^5$) and the oval (measured $d^6$) then reflects the orbital that is occupied by the extra electron that the $3d^6$ ion has in comparison to the $3d^5$ situation. This experimentally deduced shape is plotted with red dots in Figure \ref{fig:110to001}(b), and physically represents the orbital shape occupied by the sixth electron. 

Only now are we in a position to recognize that this is a cut through the [001] to [110] plane of the donut-like $d_2$ orbital, where the $d_m$ notation refers to the complex spherical harmonics $Y_2^m$. The expected shape of the cross-sectional cut through the $d_2$ orbital is shown as a grey solid line, which agrees very well with our experimental data points. To assist in visualizing the experimental cut through the total electron/hole density, we have plotted the 3D theoretical shapes (green-red shapes) of the corresponding cuts taken in the experiment (Fig. \ref{fig:110to001}). This allows one to compare the cut in the 3D electron/hole density to the 2D data obtained in the experiment, where the highlighted outlines of the cross-sectional cuts of the 3D shapes in Fig. \ref{fig:110to001} are the expected experimental curves.

To corroborate this description, and to verify that it is indeed the $d_2$ orbital that is occupied by the sixth electron, we have also obtained experimental spectra in the [100] to [010] plane (perpendicular to the Co--Co chains). Shown in Fig. \ref{fig:100to010}, we can see that the expected cross-sectional cut through the $d_2$ orbital is a horizontal cut through the donut-like shape, which is circular. Our experimental data points (green dots) obtained from integrating and summing 100.7\,eV$+102.1$\,eV peaks of this separate data set are adequately in agreement with the calculated hole density (black). As was previously described, the electron density was obtained by subtracting the hole density from a reference $d^5$ spherical hole density (pink). Again, the experimental red dots agree well with the calculated grey circular shape in Fig. \ref{fig:100to010}. Note that Figs. \ref{fig:110to001}(b) and \ref{fig:100to010} are different projections of the same hole density shape, as such, the radial scales have be plotted to be identical, and can be quantitatively compared.

We would like to note that by using the HS $3d^5$ sphere as reference, one may infer that we have effectively assumed that the $3d^6$ configuration of the Co$_{\text{trig}}$ is in the HS state. However, such an assumption is justifiable since the total effective magnetic moment ($\mu_{eff}^2 = 37 $\,$\mu_B^2$/f.u. \cite{Maignan2004}), and also the saturation magnetic moment ($M_{sat} = 4.8$\,$\mu_B$/f.u. \cite{Maignan2004}) are  larger than what an intermediate-spin or let alone a low-spin configuration could produce, taking into account that we just have firmly established that the other Co site, namely Co$_{\text{oct}}$, is non-magnetic LS $3d^6$.  But it can also be simply validated on the grounds that any \emph{ansatz} different than a HS configuration does not lead to a solution that reproduces the experimentally observed shape of the hole density.

Our finding that the $d_2$ orbital is occupied by the minority electron of the HS Co$_{\text{trig}}$ naturally explains the Ising character of the material, since this $d_2$ orbital carries a large orbital moment of 2\,$\mu_B$, and consequently generates a huge magneto-crystalline anisotropy, thereby determining the easy magnetization axis, and freezing it in the $z$-direction, i.e. along the Co--Co chain. Evidently, it is the $d_2$ orbital that is  stabilized as the lowest energy orbital by the crystal field that arises due to the trigonal prismatic coordination. 

A closer look reveals that this is in fact not trivial. Past crystal field analyses based on band structure calculations show that the $d_0$, $d_2$, and $d_{-2}$ orbitals are close in energy, but are well separated from the $d_1$, and $d_{-1}$ orbitals, as shown in Fig. \ref{fig:fitting}(b)\cite{Vidya2003,Whangbo2003,Eyert2004,Sampathkumaran2004,Takubo2005,Wu2005}. Therefore, it could have been for example the $d_0$ orbital that is occupied by the sixth electron, since its energy is very close to the $d_2$ orbital. The fact that small changes in energies leads to vastly different conclusions, perhaps at least partially explains why different types of past theoretical approaches have led to different explanations for the magnetism in Ca$_3$Co$_2$O$_6$\cite{Vidya2003,Whangbo2003,Eyert2004,Sampathkumaran2004,Takubo2005,Wu2005}. Our finding that it is the $d_2$ orbital is also special since it is extremely rare in the field of transition metal compounds that such an orbital is stabilized. It is therefore important that we now have an experiment by which we can image directly the occupied orbital without the need for calculations.

We are now able to retrospectively justify the identification and use of the three peaks in our experimental spectra required for producing the polar plots. First, we conclusively showed that the Co$_{\text{oct}}$ sites are in a 3+ LS state, where the six-fold degenerate $t_{2g}$ shell is completely filled, and the four-fold degenerate $e_g$ shell is completely empty (Fig. \ref{fig:fitting}). Thus, the only possible excitation from the $3s$ core shell to the valence $3d$ shell is to these empty $e_g$ orbitals. This is the reason why the excitations seen in the spectra that correspond to the Co$_{\text{oct}}$ site were fit with only one peak. For the Co$_{\text{trig}}$ site, we demonstrated that it is HS $3d^6$ with the sixth and minority electron occupying the $d_2$ orbital. Therefore, the $3s$ to $3d$ excited electron can reach either one of the minority orbitals in the $d_{0/-2}$ subgroup, or the $d_{+1/-1}$  subgroup. Our best results were found via the fitting procedure when the splitting (in the presence of the $3s$ core hole) between these two subgroups was 1.4\,eV. The energy separation between the peaks of the Co$_{\text{trig}}$ and Co$_{\text{oct}}$ ions reflects the chemical shift between these two sites, again, in the presence of the $3s$ core hole. Thus the existence of these peaks is retrospectively substantiated as they confirm the original hypothesis. 

{\raggedleft\textbf{Conclusion}}\\
Utilizing the substantial intensity of the dipole forbidden transitions available with s-NIXS, we are able to, in a straightforward fashion, observe the electron orbitals that are active in Ca$_3$Co$_2$O$_6$. Knowledge of these orbitals is necessary for theoretical models to more reliably calculate the various inter-site exchange interactions, and in turn to quantitatively model the magnetic structure, including the intriguing magnetization tunneling phenomena. This should also help in explaining the large and positive Seebeck coefficient reported for  Ca$_3$Co$_2$O$_6$ single crystals \cite{Mikami2003} as minority holes should dominate the electronic transport properties according to the schematic orbital diagram. Hence, the ability to experimentally observe the relevant electron orbitals without deviation to theoretical modeling is a powerful diagnosis tool for the design of novel materials. This is especially true in situations where one would like to make use of the delicate balance of competing interactions to stabilize a particular orbital state for a desired or optimized physical property. In particular, considering the large list of compounds crystallizing in structures derived from the 2H-perovskites \cite{Loye2012}, the present study should help in selecting the best candidates for quantum tunneling of the magnetization. Future investigations of this nature should open new avenues for the design of novel materials with unusual or optimized properties.

{\raggedleft\textbf{Data availability:}}\\
The data that support the findings of this study are available from the corresponding authors upon request.


{\raggedleft\textbf{Acknowledgments}}\\
B.\,L. acknowledges support from the Max Planck - University of British Columbia Center for Quantum Materials, and M.\,S., A.\,A., and A.\,S. from the German funding agency DFG under Grant No SE1441-4-1. Parts of this research were carried out at PETRA III at DESY, a member of the Helmholtz Association (HGF). We thank C. Becker, K. H\"ofer, and T. Mende from MPI-CPfS, and F.-U. Dill, S. Mayer, and H. C. Wille, from PETRA III at DESY for their skillful technical support. 

{\raggedleft\textbf{Competing Interests}}\\
 The authors declare that they have no competing financial interests.
 
{\raggedleft\textbf{Author Contributions}}\\
L.H.T., A.M., and M.W.H. initiated the project, B.L., M.S., A.A. and H.G. performed the experiment,
B.L. and L.H.T. analyzed the data, L.Z., A.C.K., and A.M. synthesized the samples, B.L., A.S., A.M., and L.H.T. wrote the 
manuscript with input from all authors.

{\raggedleft\textbf{Materials and Methods}}\\
\textbf{Experiment: }Non-resonant inelastic x-ray scattering (NIXS) measurements were performed at the High-Resolution Dynamics Beamline P01 of the PETRA-III synchrotron in Hamburg, Germany. Figures in Refs. \linecite{Yavas2019} and Ref. \linecite{Sahle2015} illustrate the experimental setup. They show the incoming beam 
($\vec{k}_i,\omega_i$), sample, scattered beam ($\vec{k}_f,\omega_f$), and the corresponding momentum transfer vector ($\vec{q}$). The energy of the x-ray photon beam incident on the sample was tuned with a Si(111) double-reflection crystal monochromator. The photons scattered from the sample were collected and energy-analyzed by an array of twelve spherically bent Si(660) crystal analyzers. The analyzers are arranged in a 3$\times$4 configuration. The energy filtering  of the analyzers ($\hbar\omega_f$) was fixed at 9.69\,keV; the energy loss NIXS spectra were measured by scanning the energy of the monochromator ($\hbar\omega_i$). Each analyzer signal was individually recorded by a position sensitive custom made LAMBDA detector. The energy calibration and resolution was ensured by pixel-wise aligning the zero energy loss feature at each measurement angle from the twelve analyzers and summing the total signal; the resolution throughout the experiment was measured to be 1.4\,eV (FWHM).

The positioning of the analyzer array determines the momentum transfer vector and the corresponding scattering triangle, which is defined by the incident and scattered photon momentum vectors, $\vec{k}$$_i$ and $\vec{k}$$_f$, respectively. The large scattering angle (2$\theta=$\,155$^{\circ}$) and use of hard x-rays chosen for this study results in a large momentum transfer. For this experiment the magnitude of our momentum transfer vector was calculated to be $\left|\vec{q}\right| = (\vec{k}_i^2+\vec{k}_f^2-2|\vec{k}_i||\vec{k}_f|\text{cos}(2\theta))^{1/2}=$ (9.6\,$\pm$\,0.1)\,\AA$^{-1}$ when averaged over all analyzers. $\vec{k}_f$ and 2$\theta$ were kept constant by fixing the energy and the position of the analyzer array. Since the energy transfer range of interest (90 to 115\,eV) is small with respect to the incident energy ($\sim$9.8\,keV), variation of $\vec{k}_i$ during energy scanning is insignificant. This guarantees that the scattering triangle was maintained throughout the course of the experiment and that $\left|\vec{q}\right|$\,$\approx$\,constant.

\textbf{Sample:}\\
The single crystal Ca$_3$Co$_2$O$_6$ sample was grown using the starting materials CaCO$_3$ and Co$_3$O$_4$, which had a molar ratio of 4:$\frac{4}{3}$; these were then mixed with K$_2$CO$_3$ in a weight ratio of 1:7. The well mixed powders were then put in an aluminum crucible and heated in air to 920$^\circ$C for 9 hours, and then left to dwell at this temperature for 24 hours. Lastly, they were slowly cooled to 850$^\circ$C at a rate of 1\,K/hr, then to 800$^\circ$C at a rate of 3\,K/hr, and finally back to room temperature in 8 more hours. Under these conditions needle  shaped crystals of several millimeters length were obtained, where the crystallographic $c$-axis lies along the length of the needle shaped sample. The orientation of the crystallographic directions was ensured to within $\pm$\,$2^{\circ}$ using a Laue diffractometer.

\textbf{Data Treatment:}\\
It is essential to the experiment that the normalization of the spectra is handled with care. A key observation is that the line shape of the Compton profile does not change with angle, as can been from Fig. \ref{fig:compton}. This is fully consistent with the fact that the scattering geometry is kept constant while rotating the sample. On the other hand, what does vary is the experimental intensity of the Compton scattering, as this is dependent on the paths the x-rays take when entering and when scattered out of the sample. That is, the incidence angle of the x-rays relative to the sample surface varies as the sample is rotated, and therefore also the interaction depth of the x-rays and the sample. However, because the theoretical Compton scattering intensity is only proportional to the total electron density in the sample---and not the scattering geometry relative to the sample surface---we should expect that this remains constant through the experiment. Therefore, the spectra in Fig. \ref{fig:compton} were normalized such that the Compton peak at 370\,eV  energy transfer was of equal intensity; and the detailed spectra (Fig. \ref{fig:stackedspectra}) in the region of the Co $M_1$ edge were likewise individually normalized using the scaling factor obtained from the Comptom profile taken at the same sample angle $\varphi$.

\clearpage

\end{document}